\documentclass[%
 reprint,
 amsmath,amssymb,aps,superscriptaddress,prl
]{revtex4-2}

\usepackage{graphicx}
\usepackage{dcolumn}
\usepackage{bm}
\usepackage{hyperref}
\hypersetup{colorlinks=true, citecolor=blue, urlcolor=black, linkcolor=blue}
\usepackage{ulem}
\usepackage{subfigure}
\usepackage{natbib}
\usepackage{xcolor} 
\definecolor{blue2}{rgb}{0,0.45,0.74}

\date{November 3, 2024}

\begin{document}
\title{Crystalline and polycrystalline regimes in a periodically sheared 2-dimensional system of disks}

\author{Siyuan Su}
\affiliation{School of Physics and Astronomy, Shanghai Jiao Tong University, Shanghai 200240, China}

\author{Jie Zhang}
\email{jiezhang2012@sjtu.edu.cn}
\affiliation{School of Physics and Astronomy, Shanghai Jiao Tong University, Shanghai 200240, China}
\affiliation{Institute of Natural Sciences, Shanghai Jiao Tong University, Shanghai 200240, China}

\author{Charles Radin}
\email{radin@math.utexas.edu}
\affiliation{Department of Mathematics, University of Texas at Austin, Austin, Texas 78712, USA}

\author{Harry L. Swinney}
\affiliation{Department of Physics and Center for Nonlinear Dynamics, University of Texas at Austin, Austin, Texas 78712, USA}

\begin{abstract}
A layer of monodisperse circular steel disks in a nearly square horizontal cell forms, for shear amplitudes SA $\le$ 0.08, hexagonal close-packed crystallites that grow and merge until a single crystal fills the container. Increasing the shear amplitude leads to another reproducible regime, 0.21 $\le$ SA $\le$ 0.27, where a few large polycrystallites grow, shrink, and rotate with shear cycling, but do not evolve into a single crystal that fills the container. These results are robust within certain ranges of applied pressure and shear frequency.
\end{abstract}

\maketitle

The compacting of rigid monodisperse particles (e.g., ball bearings) in perturbed 3-dimensional (3D) systems was studied in 1960 by Scott \cite{Scott1960}. This research revealed for certain protocols what Scott called RCP (Random Close Packing), a barrier preventing the density from rising above a volume fraction $\phi$ about 0.64. Subsequent experiments examined 3D compaction to study RCP using vertical shaking \cite{Knight1995,Richard2005} and  sedimentation \cite{Schroeter2005}. See Baule et al. \cite{Baule2018} for a history of this work.

\begin{figure} 
\centering 
\includegraphics[width=8.6cm]{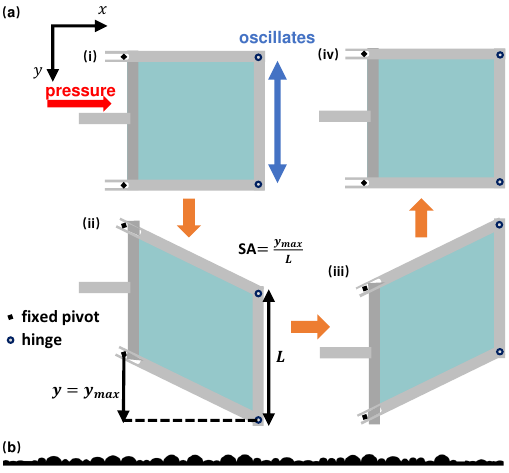} 
\caption{\label{Apparatus} 
(a) The horizontal shear cell (with sides of length $L=500$ mm) during a cycle. The shear amplitude SA$=y_{max}/L$ ranged from 0.022 to 0.264. The right lower hinge cycles in position $y$, from $y_{max}$ to $-y_{max} $ and back. The left side wall position is fixed in the $y$ direction, and the cell's top and bottom walls are connected to the side walls by hinges (black circles). The cell is under pressure 200 or 600 N/m, exerted on the left-hand wall. The shear cell is imaged at 8 frames/s with a resolution of 3 pixels/mm. In each cycle the cell is stopped for 0.5 s when the cell is rectangular (cf. Supplementary Material). (b) Crystallization near the walls is inhibited by fixing half cylinders (dia. 7 mm and 14 mm) in equal numbers in random positions on the walls.  Our analyses minimize wall effects by considering only the approximately 4200 disks further than about 35 mm from any wall.
}
\end{figure}

The present work examines compaction of a horizontal 2D system of steel disks undergoing periodic shear. Gravity and 3D effects play no role in these experiments.  Our horizontal shear cell is illustrated in Fig. \ref{Apparatus}.  

Shearing at sufficiently small shear amplitudes, SA $\le$ 0.08 (0.022 is the smallest SA examined), leads after many shear cycles to a high density single crystal. In a range of larger SA, 0.21 - 0.27, we find a second well-defined regime where a few large 2D crystallites form. With continued shearing the polycrystallites change in size and orientation, but the polycrystalline structure persists.

Our experiments are conducted using a 2D shear cell (Fig. \ref{Apparatus}(a)) containing about 5200 stainless steel disks of diameter $d=7.00 \pm 0.02$ mm and friction coefficient 0.47. The disks are sheared by oscillating the right cell wall in the $y$ direction at speed 88 mm/s. Data collected for wall speed 44 mm/s yielded the same results as the speed 88 mm/s used in the results presented in this paper. 

The results we present were obtained for initial states prepared by shearing for 100 cycles at the maximal shear amplitude, SA=0.264, with pressure P=200 N/m. The resultant initial states contain a few crystallites in different orientations. Other preparation protocols were examined and found to yield the same results as those presented here (cf. Supplemental Material). 

\begin{figure}
\centering
\subfigbottomskip=0pt
\subfigure{\label{Map-phi-a}
\includegraphics[width=8.4cm]{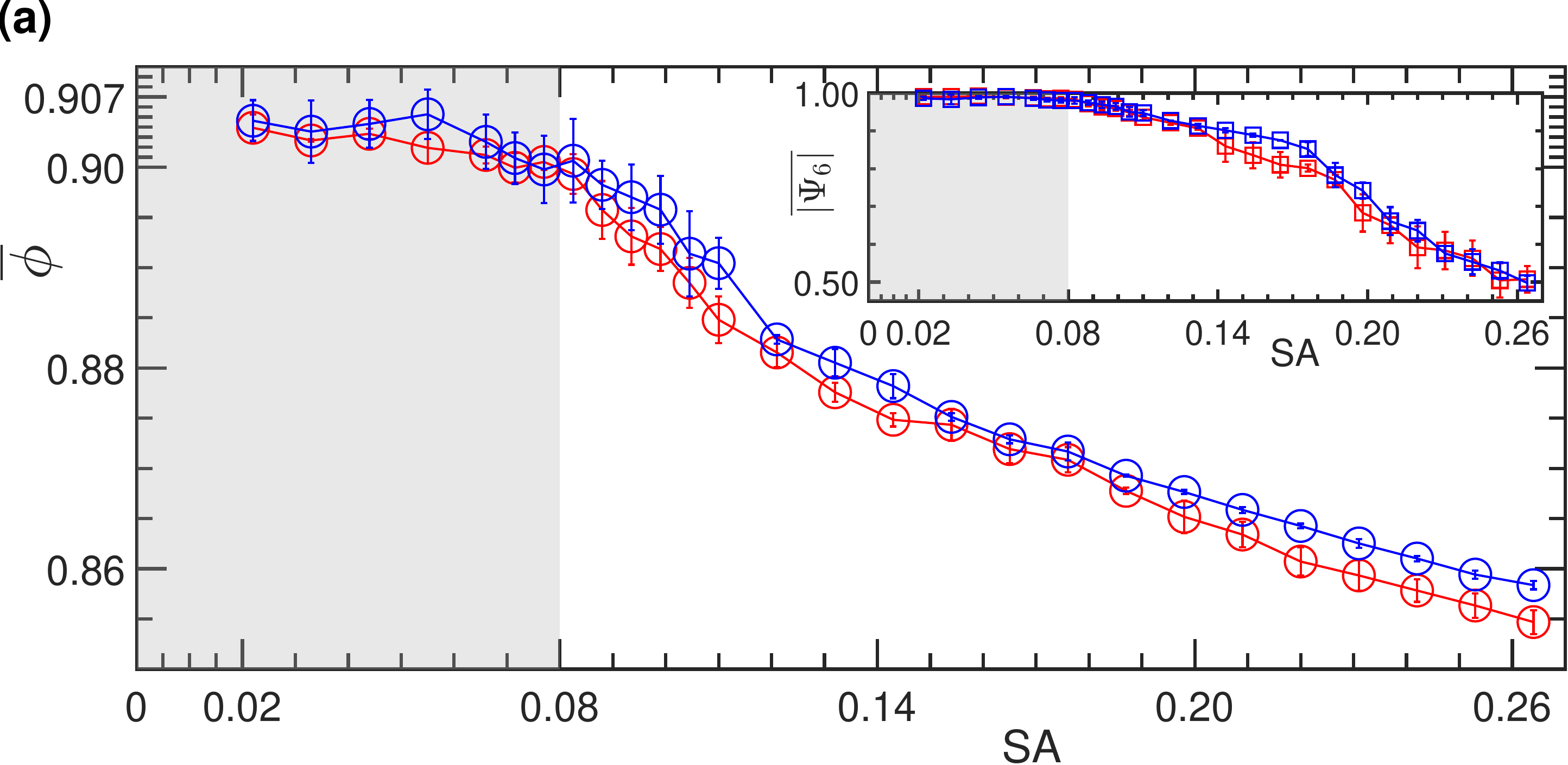}}
\subfigure{\label{Map-phi-b}
\includegraphics[width=8.4cm]{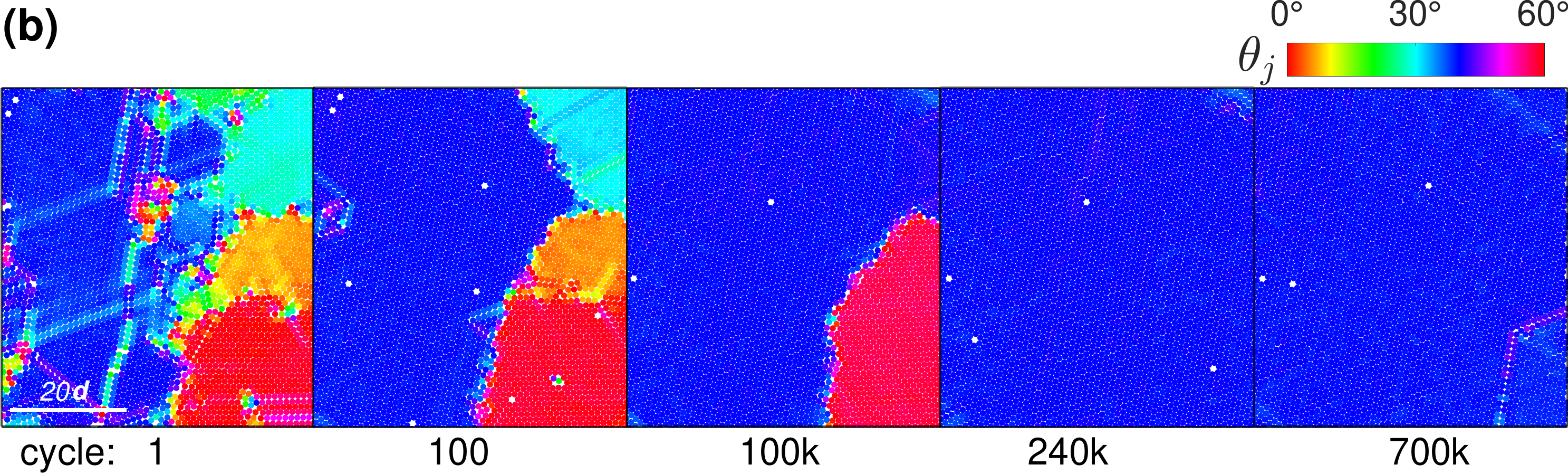}}
\caption{\label{Map-phi} 
(a) For shearing with amplitudes SA $\le$ 0.08 (shaded gray), the mean packing fraction $\overline{\phi}$  evolves until the system reaches near the close-packed limit, $\phi_{cp} \approx 0.907$. The cell is then nearly filled with a single large crystallite. The inset shows that the mean value of the global hexatic order parameter, $\overline{|\Psi_6|}$, is near unity for SA $\le$ 0.08 (see equation (\ref{global order})). (b) Cell images show evolution from an initially disordered state. The colors correspond to the angle $\theta_j$ of the local complex hexatic order parameter (see equation (\ref{local order})). The color bar shows the range of values for the local orientation angle $\theta_j$.}
\end{figure}

For small shear amplitude (SA $\le$ 0.08) the system becomes nearly perfectly ordered, and the global packing fraction $\phi$ reaches within 1$\%$ of the maximum packing of identical disks, 0.907... .  Figure \ref{Map-phi-a} shows the average result of 5 runs; the error bars are the standard deviation. The degree of order shown in the inset is obtained differently: in a cycle $t$, for each particle $j$ the local hexatic order parameter is given by the complex number
\begin{equation}\label{local order}
\psi_{6,j}(t)=\frac{1}{N_b}\sum_m^{N_b}e^{6i\Theta_{jm}}=|\psi_{6,j}|e^{6i\theta_{j}}
\end{equation}
where $\Theta_{jm}$ is the angle between the $x$-axis and the bond between particle $j$ and particle $m$ of its $N_b$ neighbors. The global order parameter for a cycle $t$ is then
\begin{equation}\label{global order}
    |\Psi_6(t)|=|\frac{1}{N}\sum_{j}^{N}\psi_{6,j}(t)|.
\end{equation}

The average $\overline{|\Psi_6|}$ of $|\Psi_6(t)|$ is obtained by averaging results over a large number of cycles, and then averaging results from different runs (cf. Supplementary Material). The values of $\overline{|\Psi_6|}$ for SA $\le$ 0.08 are close to the maximum possible value, 1, as shown in the inset of Fig. \ref{Map-phi-a}. This is possible only when all local values are aligned.  The images in Fig. \ref{Map-phi-b} show the evolution of the system with cycling from a disordered initial state.

\begin{figure}[t]
\centering
\subfigbottomskip=0pt
\subfigure{\label{Map-Psi-a}
\includegraphics[width=8.4cm]{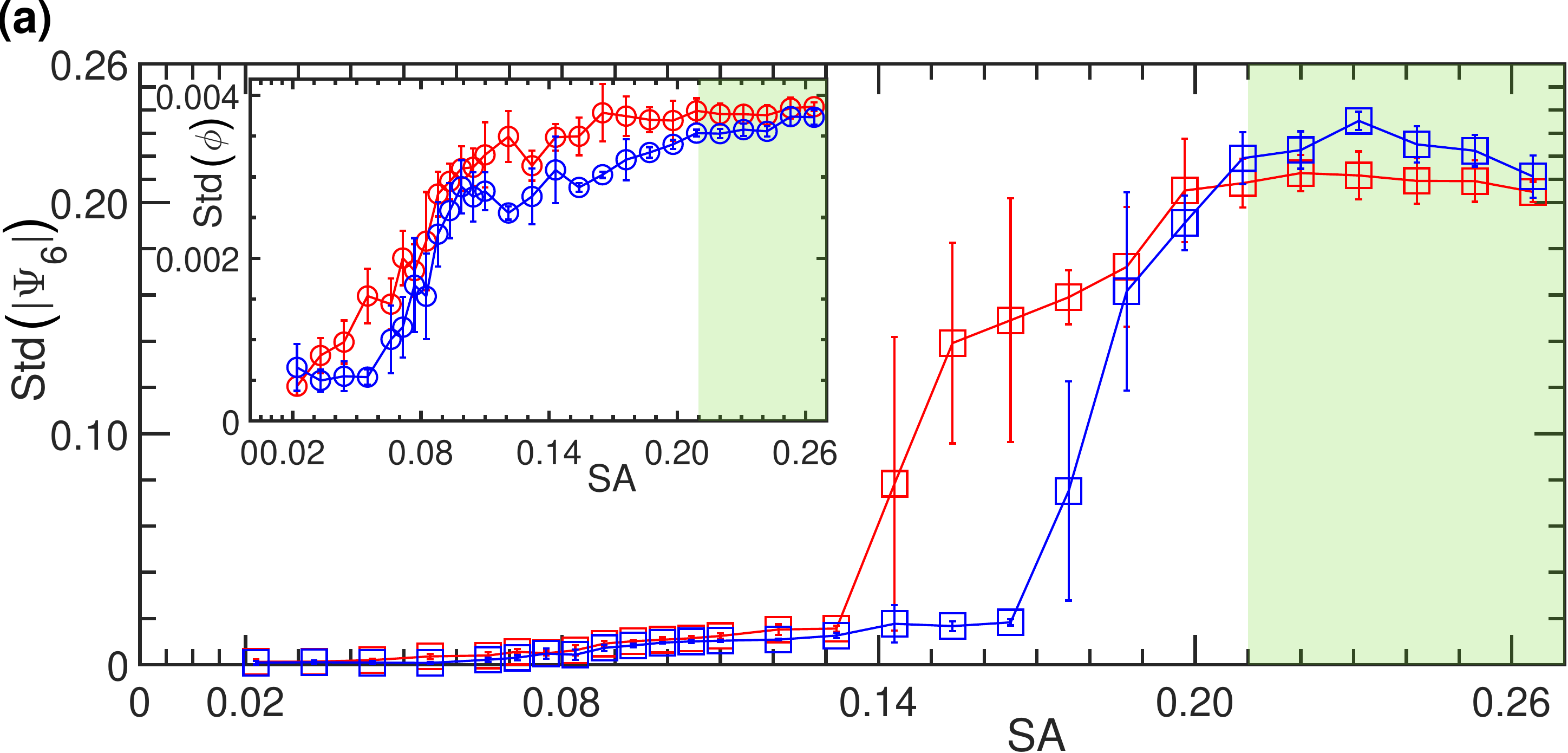}}
\subfigure{\label{Map-Psi-b}
\includegraphics[width=8.4cm]{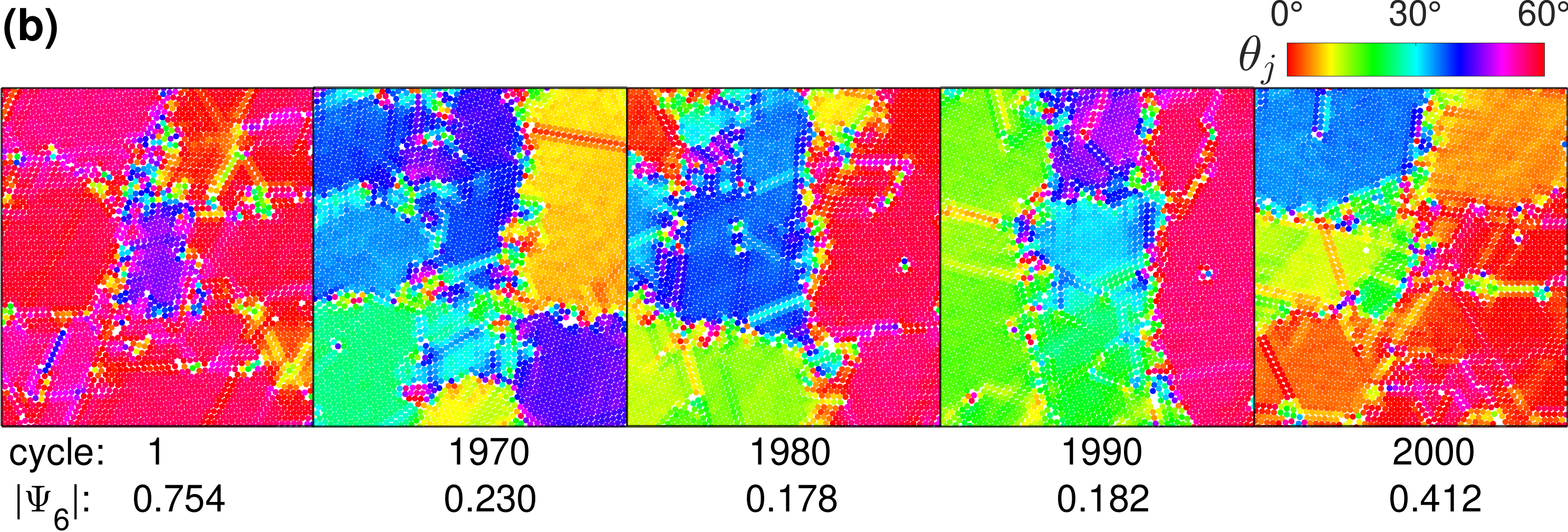}}
\caption{\label{Map-Psi} 
(a) A polycrystalline state is approached asymptotically for 0.21 $\leq$ SA $\leq$ 0.27 (shaded pale green), where the standard deviation of the global hexatic order parameter, $|\Psi_6|$ (main panel), and of the global packing fraction, $\phi$ (inset), each achieve high values with small uncertainties, for both P=200 N/m (red) and P=600 N/m (blue). (b) Snapshots from an experiment with SA=0.264 and P=200 N/m show the system evolution in the polycrystalline asymptotic state: a small number of crystals grow, shrink and change orientation (cf. color bar).}
\end{figure}

For shear cycling with shear amplitudes in a higher range, 0.21 - 0.27, a second well-defined reproducible regime emerges: a $\it {polycrystalline}$ steady-state in which the standard deviation of the global order parameter (Fig. \ref{Map-Psi-a} main figure) attains a maximum value, which is approximately constant. Also, the inset in Fig. \ref{Map-Psi-a} shows that in this range the standard deviation of the global packing fraction attains a maximum value that is approximately constant.  This regime is defined by the qualitative difference of the standard deviations in this high shear amplitude region from the corresponding standard deviations at smaller shear amplitudes. The interpretation of these states as polycrystalline is based on the local hexatic order parameter, $\psi_{6,j}$, defined in equation (\ref{local order}): $\theta_j$ is the local orientation and $6\theta_j$ is the phase angle of $\psi_{6,j}$. The snapshots in Fig. \ref{Map-Psi-b} are images from a run with SA = 0.264 (cf. Supplemental Material). These images reveal that the particles group into a small number of polycrystals that are regions of high density and well-defined orientation. As the cycling continues, these regions emerge, grow, shrink, rotate, and sometimes disappear, but the polycrystalline structure persists.

Bulk granular matter is often classified as either ``static'' or ``flowing''. The experimental research reviewed above, including Scott's experiments on RCP and experiments with cyclic shear, are usually classified as static: the materials were prepared by a variety of protocols, and then properties of the static material were analyzed.  Various methods have been used to prepare a system of particles in a range of reproducible volume fractions: combining compaction by external pressure (gravity and/or mechanical) with dilution by shaking, dilatancy (expansion) from shear, fluidization/sedimentation, and other protocols. Scott et al. (1964)\cite{Scott1964} noted that compaction and dilation together are analogous to the thermal motion of molecules in a fluid.

Ghosh et al. (2022)\cite{Ghosh2022} cyclically sheared a monolayer of 7 mm diameter acrylic circular cylinders. The particles lay on a glass plate $\it {tilted}$ from the vertical by 50 degrees. The particles were confined by another glass plate separated from the lower plate by a distance slightly larger than the particle heights. Shear was introduced by oscillating the bottom wall horizontally. The particles at the top of the cell were free. The only compression in their system was from the weight of the particles, so the system was everywhere under much less pressure than our system, and also there was a density gradient. Our results with shear amplitudes ranging from $0.02$ to $0.20$ are similar to those of Ghosh et al., but there are significant differences. Our experiment establishes well-defined asymptotic phases in two ranges of shear amplitudes: below $0.08$ and between $0.21$ and $0.27$, based on measurements of the packing fraction $\phi$ and the order parameter $|\Psi_6|$. Our system behaves differently in the shear amplitude range $0.08$ - $0.21$, while Ghosh et al. observed a sharp singularity at shear amplitude $0.065$, representing both a change in order and a change in mobility, as measured by the activity. 

Gravity plays an essential role in some phenomena observed in oscillating 3D granular media. For example, experiments by Huerta et al. \cite{Huerta2005} on a sheared homogeneous bed of grains exhibited Archimedean buoyancy, where the fluidized bed provided a gravitational counter-force remarkably similar to that felt by bodies immersed in a molecular fluid. Another example of a granular fluid exhibiting a gravitational effect is the Brazil nut effect, as studied for example by Schroeter et al. \cite{Swinney2006}. In these two experiments a height-dependent gravitational force was found to be essential but complicated. In the horizontal 2D system of this paper the absence of gravitational effects helped reveal the existence of the unexpected high shear amplitude regime. 

Our experiments are an outgrowth of experiments on the compaction of ball bearings in a 3D container by Scott et al. (1960) \cite{Scott1960}, who found a RCP barrier with a volume fraction about 0.64. In 1964 Scott et al.\ \cite{Scott1964} found that when cyclic shearing drove the system beyond RCP, there was evidence of crystallization. In 2000 Nicolas et al. \cite{Nicolas2000} used cyclic shearing of spherical particles to study the development of high density in runs with up to $10^4$ shear cycles. These measurements on shearing in the presence of gravity and an applied pressure revealed alternating compression and expansion (``dilatancy'')   \cite{Reynolds1885,Reynolds1886}). This changed the character of the research from a static picture of compaction to a dynamic picture. Nicolas et al. \cite{Nicolas2000} also found that for large cycle number there was a significant dependence on boundary conditions rather than a well-defined asymptotic state (cf. their Fig. 4). 

Crystallization in 3D is essentially different from 2D systems, which have a strong tendency to form ordered clusters, but we will revisit below the dimensionally robust subject of the alternating dominance of dilatancy and compression under cyclic shear. 

Later work examined the emergence of crystallization, for instance \cite{Mueggenburg2005,Panaitescu2012,Hanifpour2014}. Rietz et al. (2018)\cite{Rietz2018} used a laser light sheet to determine the particle positions. They found that for small shear amplitude, 0.01 rad, high density 3D configurations grew into a single large crystal, which was determined by stopping the shearing for a few times in long runs of $2\times 10^6$ cycles.  Although cyclic shear was observed to drive a granular system through RCP, there was no evidence of any stable volume fraction between RCP and the densest packing.

The present experiments on horizontal 2D arrays of circular disks have revealed two regimes with well-defined asymptotic states. In the low SA range, SA $< 0.08$, the system reaches an ordered single crystal (cf. Fig. \ref{Map-phi}) whose orientation rarely changes with cycling, while in a large SA range, $0.21-0.27$, the system forms large crystallites whose size, shape, and orientation change with the cycling, as illustrated in Fig. \ref{Map-Psi}. Each of the two well-defined regimes is a steady state rather than a static regime. For cycling at low SA, the crystalline structure is preserved, while in the high SA regime a polycrystalline structure is preserved over many cycles with the crystallites growing, shrinking, and changing orientation, and occasionally even disappearing or re-emerging.

\begin{figure}
\centering
\includegraphics[width=8.6cm]{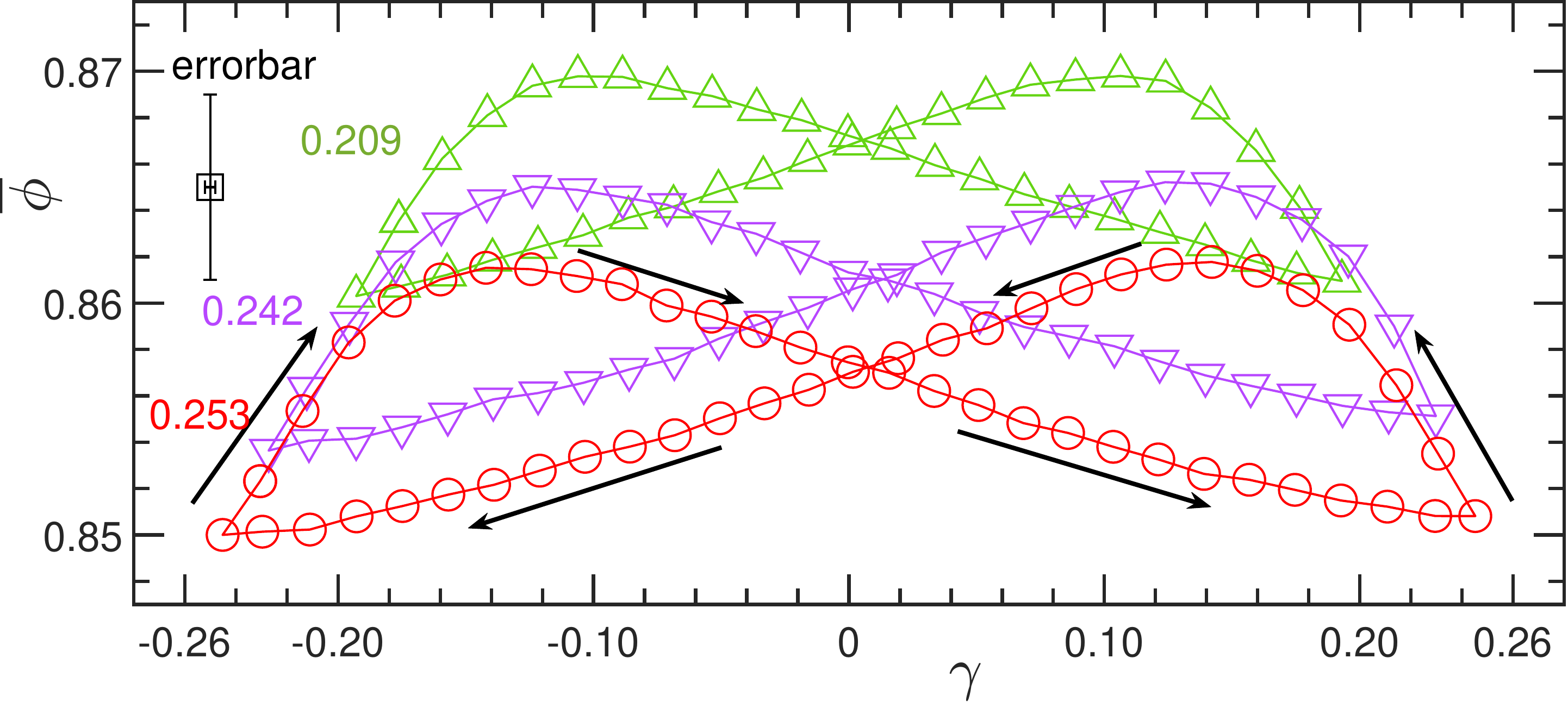}
\caption{\label{butterfly}
In the high shear amplitude regime (0.21 $ \leq$ SA $ \leq$ 0.27), the data within a cycle show that the maximum packing fraction is achieved when the cell is not rectangular. This maximum packing fraction increases smoothly with SA ($\gamma \equiv y/L$) (cf. Fig. \ref{Apparatus})), as these butterfly-shaped curves show. The black arrows show the direction of the evolution. The error bars arise from differences between different shear cycles (cf. Supplemental Material).
}
\end{figure}

Three features of our protocol may explain why we observe well-defined asymptotic states which were not seen in Nicolas et al. \cite{Nicolas2000}, where the shearing stopped and changed direction twice in each cycle. In our experiment the shearing similarly stops when changing direction. In addition, in the present experiment the shearing stops for data collection at the beginning of each cycle, and resumes without changing direction. When the shearing restarts after stopping it must overcome the equilibrium created by frictional force chains, so the addition of this third stop in each cycle may be significant. Another difference between Nicolas et al. and our experiment is that our disks lie in a horizontal 2D container, so gravity plays no role. Finally, our roughened container side walls inhibit growth of crystalline structure at those walls (see videos in the Supplemental Material), but even in the large shear amplitude regime large crystalline clusters rapidly form away from the walls, much more easily than the 3D crystallites in a 3D system.

The butterfly in Fig. \ref{butterfly} illustrates the \emph{asymptotic} behavior at high shear amplitudes. Using the conventions in Fig. \ref{Apparatus}, we define $\gamma =0$ to be the beginning of a cycle (when the shear cell is rectangular). The density variation in subsequent parts of the cycle are: a decrease to a minimum at $\gamma=$ SA, an increase to a maximum for $\gamma>0$ where the cell is non-rectangular, a decrease to a minimum  for $\gamma = -$ SA, an increase to a second maximum at $\gamma<0$, and lastly a decrease to a minimum at $\gamma = 0$.  This alternation between the dominance of shear dilation and external pressure within each asymptotic cycle is key to the effectiveness of compaction by cyclic shear. Koval et al. (2011)\cite{Koval2011} found similar butterfly-shaped curves in 3D granular material under large amplitude cyclic annular shear. Their measurements of the wall shear stress showed that after system achieved the maximum volume fraction and started to dilate, the wall shear stress approached a plateau, indicating dilatancy (cf. their Fig. 22).

\begin{figure}
\centering
\includegraphics[width=8.6cm]{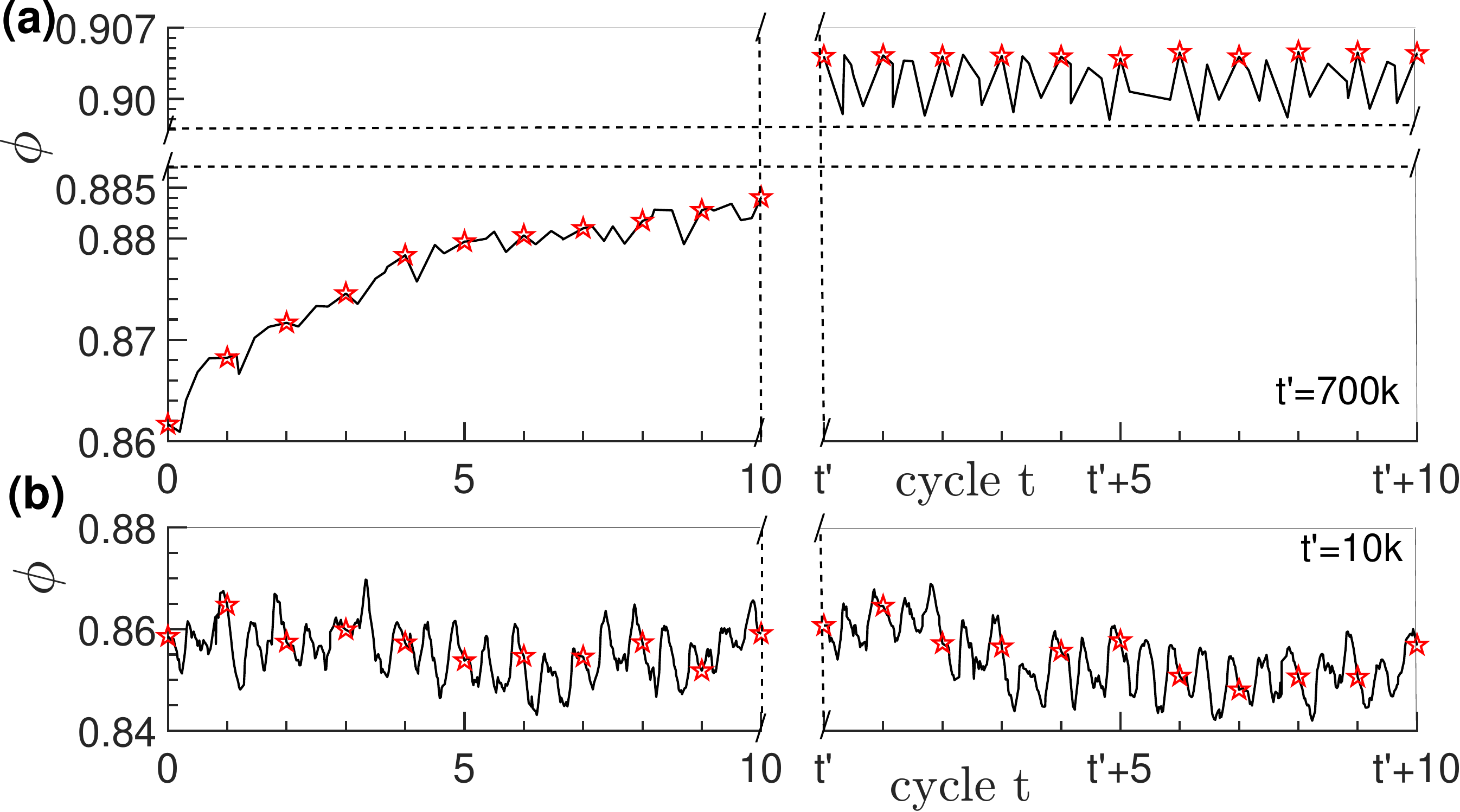}
\caption{\label{dilatancy}
(a) At low shear amplitude (here SA=0.022), a low density initial state increases in density with shear cycling. (The red stars indicate the global packing fraction values when the shear cell is rectangular at the end of each cycle.) A well-defined steady state is reached only after many cycles.  (b) At high shear amplitude (here SA=0.264), the system quickly evolves to a state with a global packing fraction about 0.86.  In this asymptotic state the density in each cycle has the complex structure shown in Fig. \ref{butterfly}. }
\end{figure}

The density evolution in the first 10 cycles in our experiment at low SA (see Fig. \ref{dilatancy}(a)) is similar in structure to that in Fig. 2 of Nicolas et al. \cite{Nicolas2000}, but our system takes much longer to achieve a steady state at low SA than at high SA. The second segment of the curve in Fig. \ref{dilatancy}(a) shows the evolution of the density for 10 cycles (after 700k cycles at low shear amplitude) shows the uniformity in the evolution for consecutive \emph{asymptotic} cycles. Rather than interpreting this evolution as a flattened butterfly, we interpret the curve as flat with small fluctuations. Figure \ref{dilatancy}(b) shows that at large shear amplitude our system achieves steady state after a very few cycles, as supported further by the similarity of the evolution in consecutive asymptotic cycles.

There are numerous advantages to working with a horizontal 2D layer of slow moving millimeter size particles, compared to a 3D system of particles in a gravitational field, both in escaping the density gradients due to gravity and in the ease of study of individual particles. Further, the addition of stopping the shearing a third time each cycle is a revealing feature of our cyclic shear protocol. We found that the net effect of competing influences produces a well-defined asymptotic crystalline state for SA $\le$ 0.08, and a well-defined asymptotic polycrystalline state for shear amplitudes between 0.21 and 0.27; no asymptotic state was found for $0.08 <$ SA $<0.21$. In future work it would be interesting to develop a method to analyze the local dynamics around each particle, to better understand the mechanisms behind compaction of grains by cyclic shear.

S.S. and J.Z. acknowledge the NSFC (No. 11974238 and No. 12274291) support and the Innovation Program of Shanghai Municipal Education Commission under No. 2021-01-07-00-02-E00138. S.S. and J.Z. also acknowledge the support from the Shanghai Jiao Tong University Student Innovation Center.

\bibliographystyle{apsrev4-1}
\bibliography{cite-paper}
\end{document}